# Polarization Controlled Ohmic to Schottky Transition at a Metal/Ferroelectric Interface

Xiaohui Liu,[1] Yong Wang,[2] J. D. Burton,[1,*] and Evgeny Y. Tsymbal[1,†]

[1] *Department of Physics and Astronomy & Nebraska Center for Materials and Nanoscience, University of Nebraska, Lincoln, Nebraska 68588-0299, USA*

[2] *Fundamental and Computational Sciences Directorate, Pacific Northwest National Laboratory, Richland, Washington 99352, USA*

Ferroelectric polar displacements have recently been observed in conducting electron-doped $BaTiO_3$. The co-existence of a ferroelectric phase and conductivity opens the door to new functionalities which may provide a unique route for novel device applications. Using first-principles methods and electrostatic modeling we explore the effect that the switchable polarization of electron-doped $BaTiO_3$ ($n$-$BaTiO_3$) has on the electronic properties of the $SrRuO_3$/$n$-$BaTiO_3$ (001) interface. Ferroelectric polarization controls the accumulation or depletion of electron charge at the interface, and the associated bending of the $n$-BTO conduction band determines the transport regime across the interface. The interface exhibits a Schottky tunnel barrier for one polarization orientation, whereas an Ohmic contact is present for the opposite polarization orientation, leading to a large change in interface resistance associated with polarization reversal. Our calculations reveal a five orders of magnitude change in the interface resistance as a result of polarization switching.

Complex oxide heterostructures exhibit an abundance of physical phenomena that involve the interplay between magnetism, electricity, and conductivity [1]. This is largely due to the presence of interfaces which have unique properties, often not existing in the bulk counterparts [2,3,4]. Especially interesting are interfaces which contain a ferroelectric material as one of the constituents. Due to their electrically switchable spontaneous polarization, ferroelectric materials are attractive for technological applications [5,6,7,8]. Interface effects induced by ferroelectric polarization involve a number of interesting phenomena [9], such as electrically controlled interface magnetization [10,11,12,13], magnetic order [14,15], magnetic anisotropy [16,17,18], in-plane [19] and perpendicular-to-the-plane transport [20,21], transport spin polarization [22,23], interface carrier density [24], and superconductivity [25]. Polarization controlled interface effects are promising for potential application in novel electronic devices.

While ferroelectric materials are normally considered as insulators, semiconducting ferroelectrics have been known for a long time [26]. The co-existence of the ferroelectric phase and conductivity is interesting because such a conducting bistable material introduces new functionalities. For example, experiments have found that the ferroelectric phase persists deeply into the metallic phase of oxygen reduced $BaTiO_{3-\delta}$ [27,28]. Theoretical studies have demonstrated that ferroelectric displacements persist up to the doping level of about $0.1e$ per unit cell (u.c.) in $BaTiO_3$ (~$10^{21}$/cm$^3$) consistent with the experimental findings [29,30]. Although in such a material an external electric field induces a flow of electric current which makes switching of the ferroelectric polarization difficult, sufficiently resistive materials may sustain the coercive voltage. For example, ferroelectric tunnel junctions are switchable despite the current flowing across them [8]. Furthermore, ferroelectric switching can be realized by an applied voltage which rises sufficiently fast in time. A recent prominent example is the resistive switching behavior of semiconducting ferroelectric $BiFeO_3$ [31]. The semiconducting behavior of $BiFeO_3$ has recently been exploited in the demonstration of polarization controlled charge transport [32,33] and switchable photovoltaic effects [34,35]. The latter behavior was qualitatively explained by a ferroelectrically driven transition from a Schottky to Ohmic contact at the interface [36]. It has also been demonstrated that a Schottky contact consisting of semiconducting $PbTiO_3$ and a high work function metal, Au, exhibits bistable conduction characteristics [37]. An on/off ratio of two orders in magnitude was observed and explained by a model in which the depletion width of the ferroelectric Schottky barrier is determined by the polarization dependence of the internal electric field at the metal/ferroelectric interface.

Driven by these developments we explore the effect of polarization on the transport regime across the interface formed between an oxide metal and a doped ferroelectric, using density-functional methods and electrostatic modeling. We predict, from first-principles, a switchable potential barrier driven by the accumulation or depletion of screening charge at the interface in response to ferroelectric polarization reversal. We demonstrate a ferroelectrically-induced change from the Ohmic transport regime, where interface conductance is metallic, to the Schottky regime, where a tunneling barrier is formed at the interface, as depicted in Fig. 1. This switching leads to a five orders of magnitude change in the interface resistance, and therefore demonstrates interesting potential for device applications.

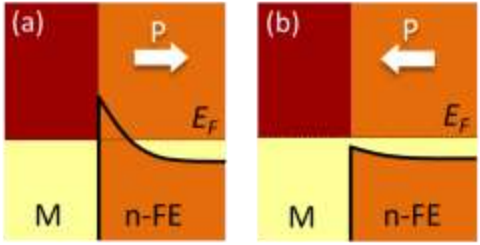

**Fig. 1**. Polarization controlled band alignment at in the interface between a metal (M) and electron-doped ferroelectric (*n*-FE). Arrows indicate the polarization direction. (a) Polarization pointing away from the interface leads to electron depletion, pulling the *n*-FE conduction band upward. (b) Polarization pointing into the interface leads to electron accumulation, pushing the *n*-FE conduction band down. In the case shown here, polarization reversal leads to a transition from a Schottky tunnel barrier (a) to an Ohmic contact (b) between M and *n*-FE.

We explore the polarization controlled contact by considering an epitaxial interface between a metallic oxide, $SrRuO_3$, and electron doped $BaTiO_3$ (*n*-$BaTiO_3$). First-principles calculations are performed using the plane-wave pseudopotential code QUANTUM ESPRESSO [38], where the exchange and correlation effects are treated within the local-density approximation (LDA). We assume that the doping of *n*-$BaTiO_3$ is 0.06 *e*/formula unit (f.u.), which is realized by the virtual crystal approximation [39] applied to the oxygen potentials in $BaTiO_3$. For this doping ($n \approx 1.9 \times 10^{21}$ cm$^{-3}$), the ferroelectric displacements remain sizable, being about 70% of those in the undoped $BaTiO_3$ [29]. The calculations are performed using periodic boundary conditions on a supercell constructed of 15.5 u.c. of $BaTiO_3$ and 10.5 u.c. $SrRuO_3$, as shown in Fig 2. We consider a $SrO/TiO_2$ interface termination at the $SrRuO_3$/*n*-$BaTiO_3$ (001) interface, which is experimentally found to be more stable, as compared to the $RuO_2/BaO$ interface [40]. We assume the same $SrO/TiO_2$ terminations at both interfaces in the supercell, which allows us to study the effect of polarization reversal at a given interface by comparing the properties of the two interfaces in the supercell for a single polarization orientation. To simulate coherent epitaxial growth on a (001)-oriented $SrTiO_3$ substrate we constrain the in-plane lattice constant of the supercell to be the calculated LDA lattice constant of cubic $SrTiO_3$, $a = 3.871$Å. Using the same approach as in the previous work [41], we perform full relaxation of the internal *z*-coordinates and overall *c*/*a* ratio of the supercell.

Fig. 2 shows the layer-resolved metal-oxygen (M-O) relative *z*-displacements across the supercell, where positive displacements indicate polarization pointing to the left. Thus, the left interface corresponds to the contact with *n*-$BaTiO_3$ polarization pointing into the $SrRuO_3$ metal, while the right interface corresponds to *n*-$BaTiO_3$ polarization pointing away from the $SrRuO_3$. In the middle of the supercell, *n*-$BaTiO_3$ exhibits bulk-like polar displacements. At the right interface, however, the M-O displacements drop sharply, while at the left interface they remain nearly constant (even slightly enhanced). This behavior is consistent with electric field profile resulting from the competition between screening, polarization charges and the built-in dipole layer at the two interfaces, as described in the electrostatic modeling discussed later.

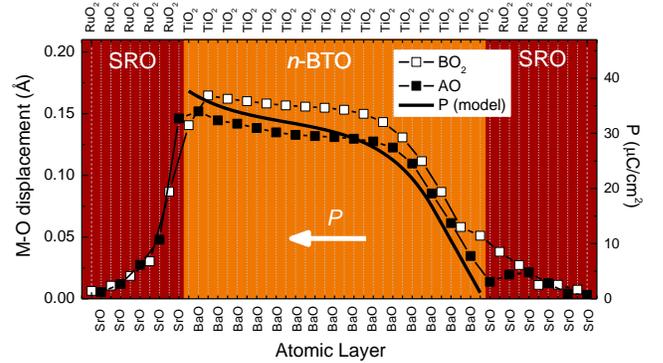

**Fig. 2.** Relative *z*-displacement between cation (M) and anion (O) on each atomic layer of the $SrRuO_3$/*n*-$BaTiO_3$ supercell. Light squares are for $BO_2$ layers (*B* = Ti or Ru) and dark squares are for *AO* layers (*A* = Ba or Sr). A positive displacement indicates that polarization is pointing to left, as shown by the arrow. The left half of the supercell corresponds to the contact with polarization pointing into the metal, while the right half of the supercell corresponds to the contact with polarization pointing out of the metal, as in Fig.1. The solid curve shows the polarization profile obtained from the electrostatic model.

Fig. 3 shows the calculated layer-resolved density of states (DOS) on the 3d-Ti orbital across the *n*-$BaTiO_3$. It is seen that at the left interface the conduction band minimum (CBM)[42] lies below the Fermi energy, implying that for polarization pointing toward the $SrRuO_3$ metal the contact is metallic (Ohmic). On the other hand, for three $TiO_2$ monolayers at the right interface the CBM lies above the Fermi energy. This implies that for polarization pointing away from the $SrRuO_3$ metal the contact exhibits a Schottky barrier. The height of this barrier is about 0.4eV and the width is about 1nm.

The major features of the CBM (Fig. 3) and polarization (Fig. 2) profiles of *n*-BTO can be captured by a continuum electrostatic model, as described in the Appendix. The effects of the $SrRuO_3$ electrodes are incorporated by interfacial boundary conditions on the *n*-BTO layer assuming (*i*) a linearized Thomas-Fermi screening length λ and relative dielectric constant ε for SrRuO$_3$ and (*ii*) an electrostatic potential step going from *n*-BTO to SrRuO$_3$, $\Delta V$, representing the built-in interface dipole, assumed to be the same at both interfaces. The polarization is modeled in the linear response regime, $P(x) = \chi\varepsilon_0 E(x) + P_0$, where $P_0$ is the polarization of bulk *n*-BTO in the absence of applied fields and $\chi$ is the linear dielectric susceptibility of the ferroelectric in response to the local electric field $E(x)$. The local electron density in *n*-BTO, $n(x)$, is determined self-consistently with the potential by incorporating the local density of states of the conduction band, which is taken from calculations of bulk *n*-BTO, only shifted by the local potential, $-e\varphi(x)$.

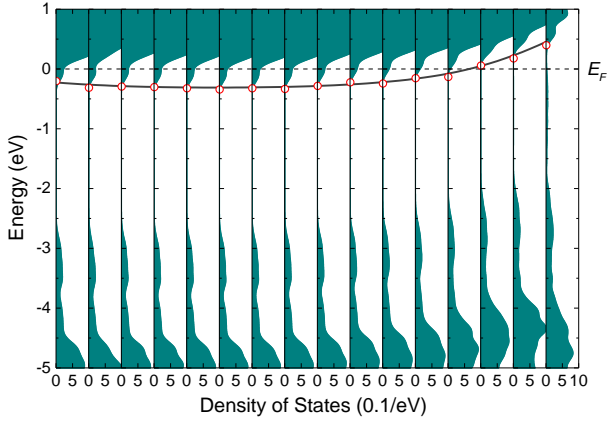

**Fig. 3.** Layer-resolved density of states (DOS) on the 3d-Ti orbital across *n*-BaTiO$_3$ (filled curves). Open circles show the conduction band minimum (CBM) obtained as described in the text. The solid curve shows the calculated CBM from the electrostatic model.

We solve the Poisson equation numerically and fit the results to the CBM profile in Fig. 3 using $\lambda$, $\varepsilon$, $\Delta V$, $\chi$ and $P_0$ as adjustable parameters. The resulting profile for the polarization and CBM are plotted alongside the first-principles results in Figs. 2 and 3, respectively, with $\lambda/\varepsilon = 0.16$ Å, $\Delta V = 0.8$ V, $\chi = 55$ and $P_0 = 32$ µC/cm$^2$ providing the best fit.

Next, we explore the electronic structure of *n*-BaTiO$_3$ in the supercell as a function of transverse wave vector $\mathbf{k}_\parallel$. In Fig. 4 we plot the $\mathbf{k}_\parallel$-resolved DOS at the Fermi energy for each TiO$_2$ atomic layer in *n*-BaTiO$_3$. Here we number the TiO$_2$ layers from 1 to 16 with layer 1 located at the left interface and layer 16 located at the right interface. The Fermi surface of bulk *n*-BaTiO$_3$ is an open tube oriented along the *z* direction, with a *z*-dependent modulation of the radius (see Fig. 8 in the Appendix). The projection of the bulk Fermi surface on the *x*-*y* plane is a slightly distorted ring, as shown in Fig. 5a, whose inner and outer radii indicate the minimal and maximal radius of the tube. When *n*-BaTiO$_3$ is placed between SrRuO$_3$ layers its Fermi surface changes. Comparing Fig. 4 to the $\mathbf{k}_\parallel$-resolved DOS for bulk *n*-BaTiO$_3$ (Fig. 5a), we see that in the middle of the supercell, e.g. for layer 7 in Fig. 4, the $\mathbf{k}_\parallel$-resolved DOS appears as a ring similar to that for bulk *n*-BaTiO$_3$. Closer to the left interface the ring is slightly distorted, but qualitatively it remains similar to the bulk one. This is due to the layer-dependent CBM remaining nearly flat at the left interface, as is evident from Fig. 3. Only for interfacial layer 1 in Fig. 4 we see a significant change in the $\mathbf{k}_\parallel$-resolved DOS which appears a disk at the $\bar{\Gamma}$ point. This feature is due to the up bending of the *n*-BaTiO$_3$ band for this interface layer (see Fig. 3) and electron density induced by the adjacent SrRuO$_3$. Thus, for polarization pointing to the SrRuO$_3$ metal layer, the contact is nearly-metallic (Ohmic) and we expect an efficient transmission across it.[43]

This behavior changes dramatically for the right interface. The upward bending of the conduction bands seen in Fig. 3 corresponds to a shrinking Fermi surface, as reflected in the reduced radius of the ring in the $\mathbf{k}_\parallel$-resolved DOS (layers 9-12 in Fig. 4) and the transformation of Fermi surface from an open tube to a closed ellipsoid (states appear at $\bar{\Gamma}$). The ring disappears at the third TiO$_2$ monolayer from the interface and the $\mathbf{k}_\parallel$-resolved DOS shows nil for layers 14-16. This is due to the CBM bending above the Fermi level. These three monolayers near the interface exhibit a gap for electron transport. Thus, for polarization pointing away from the SrRuO$_3$ metal layer, the contact is of Schottky type and we expect a reduced transmission across it.

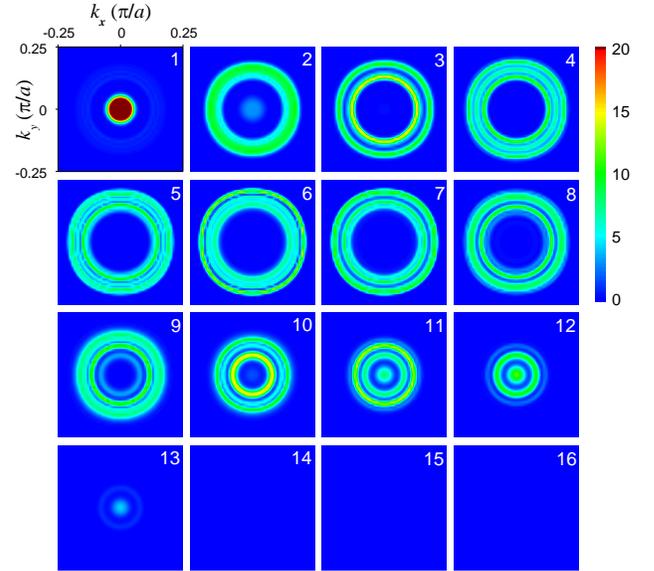

**Fig. 4.** $\mathbf{k}_\parallel$-resolved local density of states in the SrRuO$_3$/*n*-BaTiO$_3$ heterostructure, calculated at the Fermi energy for each atomic TiO$_2$ layer, numbered from the left to right interfaces corresponding to Fig. 2.

To confirm our expectations regarding the electronic transport we study the transmission across the SrRuO$_3$/*n*-BaTiO$_3$ (001) interface for two polarization orientations.

The transmission is calculated using a general scattering formalism implemented in the QUANTUM ESPRESSO [38]. In the calculation we use the left interface and the right interface in the supercell (Fig. 2) as separate scattering regions, each of which is ideally attached on one side to a semi-infinite SrRuO$_3$ electrode and on the other side to a semi-infinite $n$-BaTiO$_3$ electrode. These geometries correspond to the same SrRuO$_3$/$n$-BaTiO$_3$ junction with polarizations pointing in the opposite directions. We assume perfect periodicity in the plane parallel to the interfaces so that the in-plane component of the Bloch wave vector, $\mathbf{k}_\parallel$, is preserved for all single-electron states.

Figs. 5c and 5d show the calculated $\mathbf{k}_\parallel$-resolved transmission for polarization pointing to the SrRuO$_3$ and away from the SrRuO$_3$ respectively. The plots are limited to the region near the $\bar{\Gamma}$ point where the transmission is non-zero. This region is sampled using a uniform 51×51 $\mathbf{k}_\parallel$ mesh. The transmission distribution in the two-dimensional Brillouin zone has a similar shape for the two polarization orientations. It originates from the overlap of the Fermi surface projections of bulk $n$-BaTiO$_3$ and SrRuO$_3$ shown in Figs. 5a and 5b, respectively. The striking feature is a huge difference in the transmission magnitude for two polarization orientations. We find that polarization switching leads to a change of five orders in magnitude of transmission. The associated interface resistances[44] are $5.5\times10^2$ Ωμm$^2$ for the Ohmic contact and $3.78\times10^7$ Ωμm$^2$ for the Schottky contact.

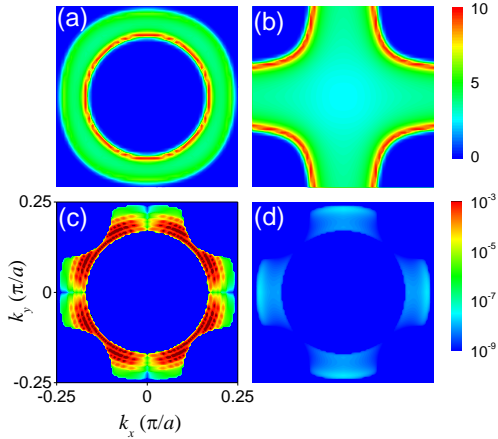

**Fig. 5.** $\mathbf{k}_\parallel$-resolved density of states at the Fermi energy in bulk $n$-BaTiO$_3$ (a) and SrRuO$_3$ (b) and ballistic transmission across the SrRuO$_3$/$n$-BaTiO$_3$ junction with polarization pointing to the SrRuO$_3$ – Ohmic contact (c), and polarization pointing away from the SrRuO$_3$ – Schottky contact (d).

## SUMMARY

We have shown that the polarization driven accumulation or depletion of free carriers at the SrRuO$_3$/$n$-BaTiO$_3$ (001) interface alters the transport regime across the interface from metallic to tunneling. We find that polarization switching leads to a five orders of magnitude change in the interface resistance. We hope that the predicted polarization controlled Ohmic to Schottky transition at the metallic oxide/doped ferroelectric interface will stimulate experimental investigations.

## ACKNOWLEDGMENTS

This research was supported by the Nanoelectronics Research Initiative (NRI) through the Center for NanoFerroic Devices (CNFD) and U.S. Department of Energy, Office of Basic Energy Sciences, Division of Materials Sciences and Engineering (DOE Grant DE-SC0004876). Computations were performed at the University of Nebraska Holland Computing Center.

## APPENDIX

### Electrostatic Model

We consider a layer of $n$-BTO from $x = 0$ to $L$ bounded on the left and right by metallic SrRuO$_3$ electrodes held in short-circuit boundary conditions. The electrodes are modeled by Thomas-Fermi screening length $\lambda$ and relative dielectric constant $\varepsilon$, and therefore the potential follows the typical form

$$\varphi_l(x) = A_l e^{x/\lambda}, \quad \varphi_r(x) = A_r e^{-(x-L)/\lambda}. \quad (1)$$

The potential inside the $n$-BTO, $\varphi(x)$, must satisfy the Poisson equation

$$\frac{\partial^2 \varphi}{\partial x^2} = -\frac{e(n_0 - n(x))}{(\chi+1)\varepsilon_0}, \quad (2)$$

where the first term on the right hand side corresponds to the uniform background density of $n$-type dopants, $n_0 = 0.06/ca^2$, and the second term corresponds to the occupied states in the conduction band.

The local carrier density, $n(x)$, is assumed to depend on $x$ only through the local potential $\varphi(x)$ and the local density of states of the conduction band, $N(E + e\varphi(x))$, where

$$N(E) = \begin{cases} 0 & E < E_c^0 \\ N_0 & E > E_c^0 \end{cases}, \quad (3)$$

$N_0$ is a constant which is determined by the nominal carrier concentration in the bulk, $n_0$, and the position of the CBM with respect to the Fermi level, $E_F$, calculated in bulk $n$-BTO from first-principles: $E_c^0 = -0.33$ eV. Therefore the average density of states is $N_0 = n_0/|E_c^0|$ and the local carrier concentration is

$$n(x) = \left(E_F - E_c^0 + e\varphi(x)\right) \begin{cases} 0 & E_c^0 - e\varphi(x) > E_F \\ N_0 & E_c^0 - e\varphi(x) < E_F \end{cases}. \quad (4)$$

Eq. (2) is subject to boundary conditions which connect $\varphi(x)$ to Eqs. (1) at $x = 0$ and $L$. $A_l$ and $A_r$ can be eliminated from these boundary conditions and the following conditions on $\varphi(x)$ emerge:

$$\varphi(0) + \Delta V + \frac{\lambda}{\varepsilon}\left((\chi+1)E(0) + \frac{P_0}{\varepsilon_0}\right) = 0$$
$$\varphi(L) + \Delta V - \frac{\lambda}{\varepsilon}\left((\chi+1)E(L) + \frac{P_0}{\varepsilon_0}\right) = 0, \quad (5)$$

where $E(x)$ is the electric field in the $n$-BTO. Note that the electrodes enter the boundary conditions only through the ratio $\lambda/\varepsilon$. Equation (2) is solved numerically subject to the boundary conditions in (5), and the CBM is related to the potential as $CBM = E_c^0 - e\varphi(x)$.

## Dependence on $\Delta V$

It is well known that the band gap calculated in density functional theory, especially in LDA, is underestimated (sometimes drastically) as compared to experiment. Therefore attempts to determine band offsets from first-principles calculations must be approached with care. In LDA we find a band gap of $E_g^{LDA} = 1.8$ eV for BaTiO$_3$, whereas in experiment it is known that $E_g^{expt} = 3.2$ eV. While little can be done for the LDA calculations to account for this issue, in our model we can make adjustments to correct for the band gap problem.

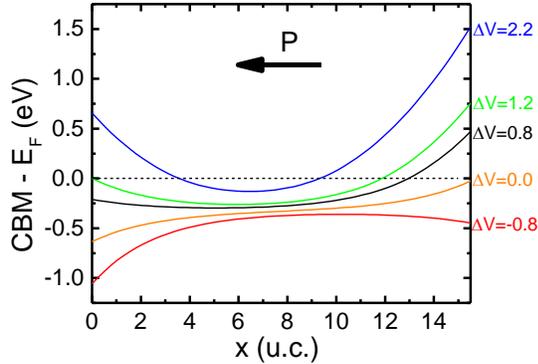

**Fig. 6.** Profile of the CBM for various interface dipoles, $\Delta V$ (in eV), but using the same best fit values found for the other parameters of the model.

The simplest correction we can make is to assume that, all else being equal, the CBM must lie higher in energy than what is predicted by LDA by a fixed difference $\Delta E_g = E_g^{expt} - E_g^{LDA} = 1.4$ eV. This correction enters our model in the interface dipole parameter, $\Delta V$. In our best fit to the LDA results we found $\Delta V = 0.8$ V, and therefore corrections for the band gap will increase this value possibly up to 0.8 V + 1.4 V = 2.2 V. In Fig. 6 we plot the CBM profile across the junction for several values of $\Delta V$, with all other parameters of the model held fixed at those of the best fit to the LDA data. Comparing the left and right interfaces, we see that for $\Delta V > 1.2$ V that there is a crossover from Ohmic to Schottky transition with polarization reversal to the interface always being of Schottky type, but with a significant difference between Schottky barrier height (SBH) and width ($w$) depending on the polarization orientation. These differences are further clarified in Fig. 7 where we plot the dependence of the interface barrier on $\Delta V$ for both polarization orientations. It is clear that the change in SBH with polarization reversal is roughly constant over a broad range of $\Delta V$, and even increases for larger $\Delta V$. Therefore we expect that our prediction of a significant change in interface resistance with polarization reversal is quite robust and independent of the deficiencies of LDA to properly predict band alignments.

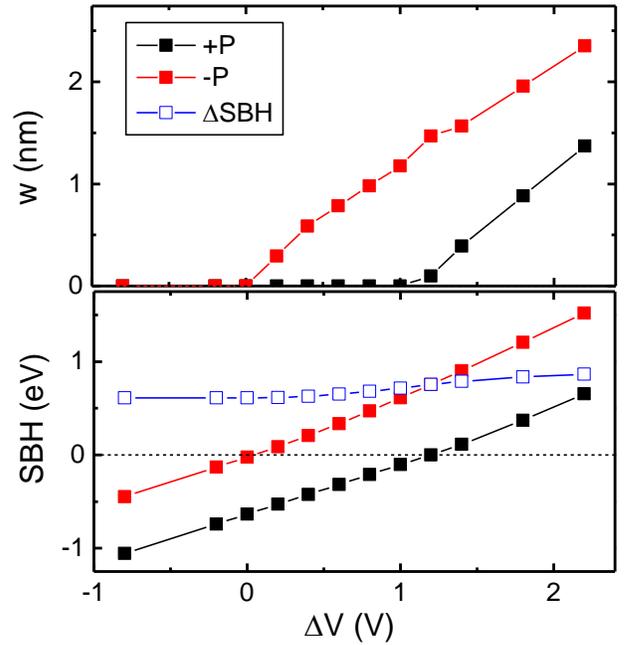

**Fig. 7.** Dependence of the Schottky barrier width ($w$) and height (SBH) on the interface dipole, $\Delta V$, for the two polarization orientations. We use the same values found for the best fit to the LDA results for the other parameters. Negative values of SBH correspond to an Ohmic contact, where $w = 0$.

## Fermi Surfaces of $n$-BaTiO$_3$ and SrRuO$_3$

One of most important considerations in the study of transport phenomena across epitaxial interfaces is the matching of Fermi surfaces. The ferroelectric displacements in $n$-BaTiO$_3$ lead to an interesting Fermi surface due to the breaking of cubic symmetry. In cubic (i.e. non-polar) BaTiO$_3$, the conduction band consists mainly of Ti $d$-states which are split by the octahedral crystal field of the oxygen cage into an upper doublet of $e_g$ states and a lower triplet of $t_{2g}$ states. The latter form the states around the conduction band minimum. The onset of polarization (i.e. off-centering of the Ti ions) leads to a splitting of the $t_{2g}$ states into an upper doublet of $d_{zx}$ and $d_{zy}$ and a lower singlet of $d_{xy}$. Therefore, when electron-doped, the free carries fill states of primarily $d_{xy}$ character. States with $d_{xy}$ character are essentially two-dimensional, with stronger coupling in the *x-y* plane than along the polarization axis, *z*. This gives rise to large band-dispersion in the plane and weak dispersion out of the plane, leading to the tube-like Fermi surface, shown in Fig. 8, for $n = 0.06$ *e*/f.u and the ring-like distributions shown in Figs. 4 and 5.

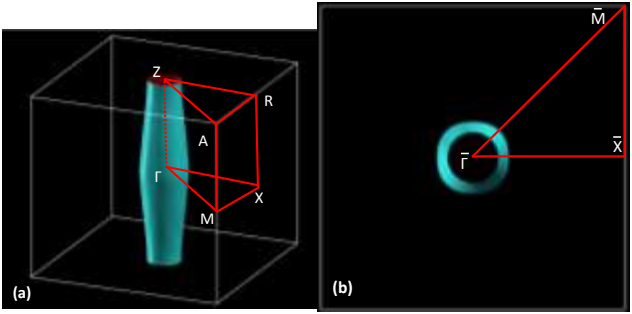

**Fig. 8.** (a) The Fermi surface of bulk *n*-BaTiO$_3$ with $n = 0.06$ *e*/f.u. Polarization and transport is along *z*. (b) View of the Fermi surface along *z* showing the origin of the ring-like distribution seen in Figs. 4 and 5.

The Fermi surface of bulk SrRuO$_3$, plotted in Fig. 9, is significantly more complicated. Given the relatively limited span of the Fermi surface of *n*-BaTiO$_3$ (Fig. 8(b)), however, the relevant features of the SrRuO$_3$ Fermi surface are limited to the cross-like region in the small range around the $\bar{\Gamma}$ point (see Fig. 9(b)). The projection of these states onto the *x-y* plane give rise to the cross features shown in Fig. 5(b), and their overlap with the Fermi surface of *n*-BaTiO$_3$ determine the shape of the transmission distributions in Fig. 5(c-d). The tetragonal structure arising from the epitaxial strain gives rise to the opening of several Fermi sheets along the *z* direction.

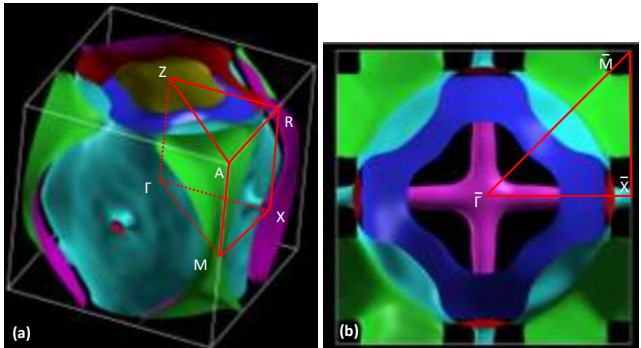

**Fig. 9.** (a) The Fermi surface of bulk SrRuO$_3$ strained in the *x-y* plane with $c/a = 1.03$, corresponding to epitaxy with an SrTiO$_3$ substrate. (b) View of the Fermi surface along *z* showing the origin of the cross-like distribution seen in Fig. 5(b).

* E-mail: jdburton1@gmail.com
† E-mail: tsymbal@unl.edu